\documentclass[prd,aps,reprint,nofootinbib,amsmath,amssymb,showpacs,showkeys,superscriptaddress]{revtex4-1}

\usepackage{graphicx}
\usepackage{natbib}
\usepackage{amsmath}
\usepackage{amsfonts}
\usepackage{xcolor}
\usepackage{bm}
\usepackage[colorlinks=true,linkcolor=blue,citecolor=blue,urlcolor=blue]{hyperref}

\newcommand{\CalN}{500}
\newcommand{\CalKSmin}{0.36}
\newcommand{\CalCovRange}{90--92}
\newcommand{\CalKSTokMin}{0.22}
\newcommand{\CalKSCnnAll}{0.002}

\newcommand{\CalMissing}{11.7}
\newcommand{\RefN}{100}
\newcommand{\RefScreenPass}{100}
\newcommand{\RefSteps}{4,000}
\newcommand{\RefBurn}{3,000}
\newcommand{\RefWalkers}{16}
\newcommand{\RefTemps}{4}
\newcommand{\RefRetained}{16,000}

\newcommand{\RefFailBadN}{3}
\newcommand{\RefStuckWalkers}{2--10}
\newcommand{\RefStuckDist}{300}

\newcommand{\RefFirstPass}{97}

\newcommand{\RefRerunTemps}{8}
\newcommand{\RefRerunSteps}{20,000}

\newcommand{\RefContSteps}{20,000}

\newcommand{\ErrRatio}{1.17--1.30}

\newcommand{\TruthRefCov}{82--87}

\newcommand{\TruthZRef}{1.14--1.18}

\newcommand{\WidthRange}{1.39--1.72}

\newcommand{\KLfm}{8.43}

\newcommand{\KLtok}{12.08}

\newcommand{\KLcnn}{28.37}

\newcommand{\KLis}{0.012}

\newcommand{\KLfloor}{0.023}

\newcommand{\IsN}{100}
\newcommand{\IsESSmed}{695}

\newcommand{\IsEffMed}{14}

\newcommand{\TimeFM}{8.0}

\newcommand{\TimeISwall}{48}

\newcommand{\TimePTwall}{129}
\newcommand{\TimePTconvWall}{166}

\newcommand{\TrainHours}{28}

\newcommand{\ExGap}{5.3}

\begin{document}

\title{Wavelet-domain flow-matching posterior estimation for LISA signals with data gaps}

\author{Ruiting Mao} \affiliation{Department of Statistics, University of Auckland, Auckland 1010, New Zealand}
\author{Jeong Eun Lee} \affiliation{Department of Statistics, University of Auckland, Auckland 1010, New Zealand}
\author{Matthew C. Edwards} \affiliation{Department of Statistics, University of Auckland, Auckland 1010, New Zealand}


\begin{abstract}
The Laser Interferometer Space Antenna (LISA) data stream will contain scheduled and unscheduled gaps, which break the stationarity of the instrumental noise and invalidate the standard frequency-domain likelihood. Amortized simulation-based inference (SBI) avoids explicit likelihood evaluations, but its application to months-long observations requires compressing high-dimensional, gapped time series while preserving information about source parameters. We develop an end-to-end amortized posterior estimation pipeline: the gapped data are mapped to noise-scaled Wilson--Daubechies--Meyer (WDM) wavelet coefficients and summarized by a convolutional--Transformer network, trained jointly with a flow-matching (FM) posterior estimator, with optional importance-sampling correction. For a LISA-like signal observed for 3 months, the amortized FM posteriors are well calibrated on simulated observations whose waveforms and gaps were not seen in training. The FM posteriors agree closely with the benchmark, the posteriors from parallel-tempered Markov chain Monte Carlo (PTMCMC), although they exhibit moderately greater dispersion. Importance sampling efficiently reweights the FM posterior, achieving close agreement with the benchmark at modest additional cost.
\end{abstract}

\maketitle

\section{Introduction} \label{sec:introduction}

The Laser Interferometer Space Antenna (LISA)~\cite{lisa:2017, colpi2024lisadefinitionstudyreport, baker2019laser}, adopted by the European Space Agency in 2024, will be the first space-based observatory of gravitational waves, in the millihertz band ($0.1$\,mHz--$1$\,Hz). Ground-based detectors such as LIGO, Virgo and KAGRA observe transients lasting from milliseconds to minutes; LISA will instead be dominated by sources that remain in band for months or years~\cite{Burke:2025}. Its data stream will hold many sources overlapping in time and frequency: massive black hole binaries~\cite{berti2006gravitational, sesana2005gravitational}, extreme mass-ratio inspirals~\cite{Gair_2017, Babak_2017} and a foreground of Galactic binaries~\cite{Cornish_2017, Littenberg_2019, Willems_2008}. Consequently, the extraction of scientific information necessitates a ``global fit'' analysis to simultaneously resolve these entangled signals~\cite{littenberg:2023}, a task that is particularly sensitive to the spectral leakage and nonstationarities introduced by unavoidable data gaps and glitches~\cite{Baghi:2019, pearson2025handling, Burke:2025, Castelli_2025}.

Over such durations the data will not be continuous. As LISA Pathfinder showed~\cite{Armano:2016, Armano:2018}, the stream will be interrupted by scheduled maintenance, such as antenna repointing and mechanism adjustments, and by unscheduled events such as micrometeoroid impacts and instrumental glitches~\cite{Burke:2025, pearson2025handling}. In standard gravitational-wave inference, analysis is typically conducted in the frequency domain under the assumption of stationary noise, which permits the computationally efficient diagonal approximation of the noise covariance matrix~\cite{pearson2025handling}. However, gaps break the time-translation invariance of the noise and cause spectral leakage~\cite{Baghi:2019, Castelli_2025}, so the Whittle likelihood applied across the whole observation becomes statistically inconsistent~\cite{Burke:2025, wang2024windowinpaintingdealingdata}. If unaddressed, these artifacts could lead to biases in astrophysical parameter estimation and losses of signal-to-noise ratio (SNR)~\cite{gair2025contamination}, as documented for Galactic binaries~\cite{carre2010effect} and for massive black hole binaries, particularly when a gap falls near the merger~\cite{Dey_2021, Mao_2025}.

Simple interpolation or mean imputation cannot restore the phase-coherent structure of gravitational-wave signals~\cite{Blackman_2014}, so the usual remedy has been apodization: smooth windows that suppress the leakage~\cite{carre2010effect, Dey_2021, wang2024windowinpaintingdealingdata}. Burke \textit{et al.}~\cite{Burke:2025} showed, however, that windowing alters the noise itself: it violates the stationarity (Toeplitz) condition behind the Whittle likelihood and can bias the parameters, and smoother tapers reduce the bias only by discarding the information in the tapered data. Bayesian data augmentation~\cite{Baghi:2019} instead samples the missing data alongside the parameters; it is statistically consistent but, because of dense matrix operations on the conditional distributions, prohibitively expensive for long signals. Pearson and Cornish~\cite{pearson2025handling} made augmentation cheaper in the wavelet domain, assuming the noise is locally stationary within each time--frequency pixel. Our previous work imputed the gaps with a neural network before MCMC~\cite{Mao_2025}, which requires denoising the data first. The exact time-domain likelihood avoids leakage but requires a dense $N\times N$ noise covariance, with an $\mathcal{O}(N^3)$ cost for $N\sim10^5$--$10^6$~\cite{Burke:2025, pearson2025handling}; short-time Fourier methods restrict the transforms to contiguous segments and trade time against frequency resolution~\cite{stft_paper1, stft_paper2}.

Posterior estimation in amortized simulation-based inference (SBI)~\cite{cranmer2020frontier} sidesteps likelihood evaluations: a network learns the posterior from simulations, so any effect that can be simulated, such as gaps and changes in the noise, enters through the training data rather than through a model of the likelihood. Conditional variational autoencoders~\cite{gabbard2022bayesian} and normalizing flows~\cite{green2020gravitational, Dax2021, Williams2021} give rapid, amortized parameter estimation for ground-based detectors~\cite{liang2025recent}, and learned estimators have handled drifting noise spectra~\cite{Wildberger2023}, real detector noise~\cite{Raymond2025}, non-Gaussian noise~\cite{legin2025gravitational} and glitches~\cite{Xiong2025}. For LISA, SBI has been applied to massive black hole binaries~\cite{Liang_2024, Spadaro2026}, extreme mass-ratio inspirals~\cite{Liang_2025} and Galactic binaries~\cite{srinivasan2025, Delmond2026}, and proposed for a stochastic background observed with time-varying instrumental noise~\cite{Alvey2025}; the recent analyses of individual sources assume continuous data and leave gaps to future work~\cite{Spadaro2026, Delmond2026}. For the amortized posterior estimator, we use flow matching (FM)~\cite{lipman2022flow, wildberger2023flow}, which learns a velocity field that carries Gaussian draws to the posterior. It is trained by regression onto straight paths between the two, without integrating the flow, and posterior samples are drawn by integrating it as an ordinary differential equation. Unlike discrete normalizing flows, it places no constraint on the network architecture and still gives the exact density of its posterior estimate; in gravitational-wave inference it was more accurate than comparable discrete flows at a lower training cost~\cite{wildberger2023flow}, and it has been trained on real detector noise~\cite{Raymond2025}. The exact density also lets the FM posterior distribution (FMPD) serve as a proposal for importance sampling (IS) with the likelihood~\cite{dax2023neural, Spadaro2026}, which we add as an optional correction.

For multi-month LISA data, the difficulty of amortized posterior estimation shifts from likelihood evaluation to the compression of the data: the network must compress more than a million gapped samples into a short summary while retaining information about the source. Current amortized posterior analyses of LISA sources start from the Fourier coefficients of continuous data. Delmond \textit{et al.}~\cite{Delmond2026} pass a narrow frequency band directly to the flow and expect an embedding network to be needed for wider bands, while DINGO~\cite{Dax2021}, now extended to LISA massive black hole binaries~\cite{Spadaro2026}, compresses the data with a fully connected residual network whose first layer is seeded with principal components of clean waveforms. Gaps break the phase coherence of the signal and make the Fourier representation less suitable~\cite{Delmond2026}, and a multi-month series cannot be Fourier transformed as a whole without spreading the gap artifacts. The discrete Wilson--Daubechies--Meyer (WDM) wavelet transform~\cite{Cornish:2020:PhRvD, pearson2025handling} instead gives an orthogonal basis in which each gap affects only the time--frequency pixels that overlap it. Convolutional residual networks have conditioned normalizing flows on spectrograms of ground-based data~\cite{lanchares2025}, and Transformer encoders~\cite{vaswani2017attention} capture non-local dependencies and flexible sequence structure in gravitational-wave inference~\cite{kofler2025flexible, leyde2026endtoendpopulationinferencegravitationalwave}. For gapped LISA data, convolutional tokenization can extract localized frequency tracks and self-attention can relate segments separated by gaps. Keeping the signed wavelet coefficients, rather than their magnitudes, retains the phase and sign information of the signal.

In this work we present an end-to-end inference pipeline for long-duration LISA signals with data gaps. It couples a summarizer, consisting of the WDM representation with noise normalization and a convolutional-tokenized Transformer encoder, with an FM posterior estimator, and we test it against likelihood-based sampling. Where the likelihood can be evaluated, an optional importance-sampling step corrects the FMPD. Our contributions are:
\begin{itemize}
\item \textbf{Amortized WDM-domain posterior estimation for data with gaps.} The data stream summarizer and the FMPD estimator are trained together on simulated LISA data streams interrupted by gaps. Signed, noise-normalized WDM coefficients preserve both the amplitude and the phase-coherent structure of the signal, without imputation or an $\mathcal{O}(N^3)$ inversion of the time-domain noise covariance, and the amortized FM posteriors are calibrated over \CalN{} simulated observations.
\item \textbf{Convolutional--Transformer summarizer and its empirical comparison.} We introduce an asymmetric convolutional-tokenized Transformer summarizer that compresses long observations into a sequence of tokens while preserving fine frequency resolution. It maps any gapped WDM image to a context vector, so it is not tied to FM or to this signal: it could condition other neural posterior or likelihood estimators, or be retrained for other LISA sources and gap models (Sec.~\ref{sec:discussion}). We further compare it with two alternative summarizers, token pooling and a globally pooled convolutional network (pooled CNN), each trained jointly with an FMPD estimator of the same architecture, on the same data and with the same schedule. The convolutional--Transformer summarizer gives the posteriors closest to those of parallel-tempered MCMC (PTMCMC), demonstrating its effectiveness in preserving information relevant to posterior inference; the pooled network also misrepresents the correlation of $f_0$ and $\dot f_0$. However, the posteriors from all three summarizers are wider than PTMCMC, and the FMPD means lie further from the truth on average than the PTMCMC means (Appendix~\ref{app:truth}), so some information is still lost.

\item \textbf{On-the-fly simulation and augmentation.} Only clean waveforms are stored; during training each is combined with a fresh noise realization and a sampled gap mask and transformed to the WDM domain within the training step. This avoids storing a static library of wavelet maps for long observations and exposes the network to new noise realizations, and new combinations of noise and gaps, throughout training rather than to a fixed set.
\item \textbf{Optional posterior estimate correction using IS.} With the FMPD as proposal, likelihood reweighting~\cite{dax2023neural} combines the speed of amortized inference with likelihood-based accuracy: as the number of draws grows, the reweighted posterior converges to the posterior under the likelihood. It corrects FMPDs in an average \TimeISwall\,s per observation, yielding posterior estimates comparable to those obtained using PTMCMC.
\end{itemize}

As a test case, the pipeline is applied to a LISA-like chirping sinusoid observed for three months, with three free parameters. We evaluate the learned posteriors in two ways, on new simulated observations whose signals and gaps were not seen in training: by their calibration, and by their agreement with likelihood-based posteriors from PTMCMC. Section~\ref{sec:preliminaries} specifies the problem and the simulations. Section~\ref{sec:pipeline} describes the pipeline. Section~\ref{sec:results} presents the calibration of the posteriors and their agreement with PTMCMC, including comparisons of summarizers and the importance-sampling correction for the FMPDs. Section~\ref{sec:discussion} discusses the limitations and possible extensions. The appendices include the network details and supplementary results.

\section{Problem and simulations} \label{sec:preliminaries}

\subsection{Data model} \label{subsec:problem}

We consider the time-delay interferometry (TDI) $A$ channel, sampled at $\Delta t=5\,\text{s}$ for three-month signals ($N=1{,}572{,}864$). The gapped data are
\begin{equation}\label{eq:data_model}
\hat{d}(t)=m(t)\odot\left[h(t;\boldsymbol{\theta})+n(t)\right],
\end{equation}
where $h(t;\boldsymbol{\theta})$ is the detector response to a source with parameters $\boldsymbol{\theta}$, $n(t)$ is stationary Gaussian noise with one-sided power spectral density (PSD) $S_n(f)$, $\odot$ is the element-wise product, and the mask $m(t)\in\{0,1\}$ is zero where data are missing. Since the noise is continuous, an observed sample is almost never exactly zero, so the networks can identify gaps from $\hat{d}$ itself.

\subsection{Signal} \label{subsec:signal}

To isolate the treatment of gaps we consider one source and leave overlapping sources to global-fit analyses~\cite{littenberg:2023, srinivasan2025}. The source is a chirping sinusoid with the polarization structure of a Galactic binary~\cite{PhysRevD.106.103001},
\begin{align}
h_{+} &= -A\,(1+\cos^2\iota)\cos\Phi(t), \nonumber\\
h_{\times} &= -2A\cos\iota\,\sin\Phi(t), \nonumber\\
\Phi(t) &= -\phi_0 + 2\pi\left(f_0 t + \tfrac{1}{2}\dot{f}_0 t^2 + \tfrac{1}{6}\ddot{f}_0 t^3\right), \label{eq:waveform}
\end{align}
with amplitude $A$, inclination $\iota$, initial phase $\phi_0$, initial frequency $f_0$, frequency derivative $\dot{f}_0$ and second frequency derivative $\ddot{f}_0 = 11\dot{f}_0^2/(3f_0)$. The polarizations are rotated by the polarization angle $\psi$ and projected onto the $A$ channel with \texttt{fastlisaresponse}~\cite{fastlisaresponse_url}, for equal-arm analytic orbits and first-generation TDI~\cite{Tinto_TDI}. We infer $\boldsymbol{\theta} = (\log_{10}A,\ \ln f_0,\ \ln\dot{f}_0)$ and fix $\iota$, $\phi_0$, $\psi$ and the ecliptic latitude $\beta$ and longitude $\lambda$ at the values of Table~\ref{tab:setup}. The parameters are chosen as a demanding test of the pipeline, not as a typical Galactic binary. With $\dot{f}_0\sim10^{-10}\,\text{Hz/s}$ the frequency rises from 2.0 to about $3.35\,\text{mHz}$ during the observation, so that the track crosses 28 WDM frequency rows and the network must follow it across the gaps; a detached double white dwarf at 2\,mHz, with $\dot{f}_0\sim10^{-17}$\,Hz/s, would stay within one row. For a binary driven by gravitational radiation these values would imply coalescence within about 87 days, where Eq.~\eqref{eq:waveform} is not an accurate waveform, so we treat $h$ as a controlled test signal and draw no astrophysical conclusions. The prior is uniform in $\boldsymbol{\theta}$ over a box centred on $\big(\log_{10}(1.5\times10^{-21}),\ \ln(2\,\text{mHz}),\ \ln(10^{-10}\,\text{Hz/s})\big)$ with half-widths $(0.1,\,10^{-4},\,10^{-4})$. In $f_0$ this is only $\pm2\times10^{-7}\,\text{Hz}$, about three Fourier bins, but several hundred times the posterior standard deviation, so the posteriors are dominated by the data. The prior is nevertheless local: the network would operate after a search stage that has localized the source. We precompute 30,000 clean waveforms from this prior, 24,000 for training and 6,000 for validation.

\subsection{Noise and gaps} \label{subsec:noise}

The noise is stationary and Gaussian with the $A$-channel PSD of \texttt{LISA Analysis Tools}~\cite{katz_michael_2025_17138723}; each realization is drawn in the frequency domain and transformed to the time domain, and the noise is added before masking. LISA data will contain scheduled gaps, for operations such as antenna repointing, and unscheduled ones, from events such as instrument outages and micrometeoroid impacts~\cite{Dey_2021, Burke:2025}. We generate masks with \texttt{lisaglitch}~\cite{lisaglitch}, with scheduled gaps of about 100\,s three times a day, of 3.3\,h every two weeks and of 40\,h six times a year, and unscheduled gaps of 24\,h, 60\,h and 66\,h, on average 30, 3 and 4 times a year. Training uses a library of 1,000 such masks, in which the missing fraction ranges from 3.1\% to 16.5\%, with a median of 11.9\%; since each mask is paired with fresh noise at every training step, the data stay varied while the masks are generated only once. The calibration set (Sec.~\ref{subsec:calibration}) uses masks generated in the same way, but none of which appears in training.

\subsection{Why use amortized SBI} \label{subsec:sbi}

For stationary noise the likelihood is evaluated in the frequency domain with the Whittle approximation~\cite{finn1992detection, Flanagan:1997kp},
\begin{align}
\ln p(d \mid \boldsymbol{\theta}) = -\tfrac{1}{2}\,(d - h \,|\, d - h) + \text{const} \nonumber
\end{align}
where $(a|b) = 4\,\text{Re}\int_0^\infty \tilde{a}(f)\tilde{b}^{*}(f)/S_n(f)\,\mathrm{d}f$. This relies on uncorrelated frequency bins. With gaps, the transform of $m(t)\odot n(t)$ is the convolution of $\tilde{n}(f)$ with the transform of the mask, which correlates the bins~\cite{Burke:2025, pearson2025handling}; the exact likelihood then requires the dense $N\times N$ covariance of the gapped noise.

 Amortized SBI replaces the likelihood evaluation by simulations~\cite{cranmer2020frontier, Dax2021}: parameters are drawn from the prior, observations are simulated with Eq.~\eqref{eq:data_model}, and a network is trained to approximate the posterior $p(\boldsymbol{\theta} \mid \hat{d})$, with its approximation denoted by $q(\boldsymbol{\theta} \mid \hat{d})$. Our network has two parts (Fig.~\ref{fig:architecture}): a summarizer $c_\psi$ that compresses $\hat{d}\in\mathbb{R}^N$ into a context vector, and an FMPD estimator that turns a standard Gaussian into the posterior conditioned on that context. Both are trained together with one objective, and once trained they produce posterior samples for any observation within the prior without evaluating a likelihood.

\section{Inference pipeline} \label{sec:pipeline}

\begin{figure*}[t]
  \centering
  \includegraphics[width=\textwidth]{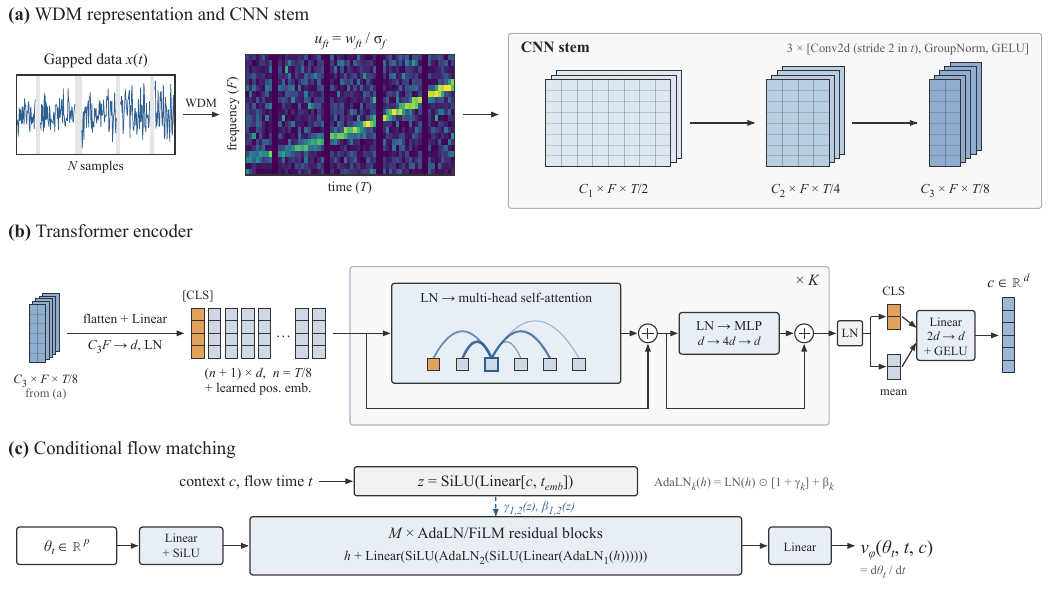}
  \caption{The pipeline. (a)~The corrupted data are mapped to a noise-scaled WDM time--frequency image, from which a convolutional stem extracts local features. (b)~A Transformer summarizer compresses the resulting sequence of tokens into a context vector $c$. (c)~A conditional flow-matching network $v_\varphi(\boldsymbol{\theta}_t, t, c)$ transports standard-normal samples to posterior samples. The data shown are schematic.}
  \label{fig:architecture}
\end{figure*}

The pipeline maps the corrupted data to a noise-scaled time--frequency image, summarizes the image into a context vector, and conditions a flow-matching posterior estimator on it (Fig.~\ref{fig:architecture}).

\subsection{Wavelet representation and noise scaling} \label{subsec:wdm_preprocessing}

The discrete WDM transform is a complete, orthonormal, real-valued tiling of the time--frequency plane~\cite{necula2012transient, Cornish:2020:PhRvD}. A series of $N = N_f N_t$ samples is mapped onto $N_f$ frequency rows and $N_t$ time bins, with pixels of duration $\Delta T = N_f \Delta t$ and bandwidth $\Delta F = 1/(2\Delta T)$:
\begin{equation}
w_{nm} = \sum_{k=0}^{N-1} \hat{d}(t_k)\,\chi_{nm}(t_k), \label{eq:wdm_transform}
\end{equation}
where the basis function $\chi_{nm}$ is a Meyer window centred at time $n\Delta T$, modulated by $\sqrt{2}\cos(\pi m t/\Delta T)$ or $\sqrt{2}\sin(\pi m t/\Delta T)$ depending on the parity of $n+m$~\cite{Cornish:2020:PhRvD}. We compute it with the fast transform of \texttt{pywavelet}~\cite{pywavelet}. Because each basis function is localized in time, a gap affects only the coefficients whose support overlaps it, whereas in a Fourier transform it affects every coefficient~\cite{pearson2025handling}; for stationary noise the coefficients are approximately uncorrelated with a variance set by the PSD at the row frequency~\cite{Cornish:2020:PhRvD}. We use $N_f = 2048$ and $N_t = 768$, giving $\Delta T\approx2.8\,\text{h}$ and $\Delta F\approx48.8\,\mu\text{Hz}$, and keep the 82 rows whose central frequencies lie in $[1, 5)\,\text{mHz}$, which contain the track for every draw from the prior. The resulting image $W\in\mathbb{R}^{82\times768}$ has 25 times fewer entries than $\hat{d}$.

The LISA noise varies strongly with frequency, so we divide each row by its noise level,
\begin{equation} \label{eq:noise_normalization}
u_{ft} = w_{ft}/\sigma_f,
\end{equation}
where $\sigma_f$ is the standard deviation of noise-only, gap-corrupted WDM coefficients in row $f$, estimated once from independent noise realizations with masks from the library, in double precision (Appendix~\ref{app:training_reproducibility}), and then held fixed. This equalizes the noise level across rows; it is a rescaling, not a whitening of the gap-modified noise covariance. It also keeps the sign of every coefficient. Since the basis functions alternate between cosine and sine modulation, the signs carry the phase of the signal, and the phase evolution is what constrains $f_0$ and $\dot{f}_0$ most tightly; signed coefficients let the network use phase coherence across the segments on either side of a gap.

\begin{table}[t]
\caption{Simulation and training setup.}\label{tab:setup}
\begin{ruledtabular}
\begin{tabular}{ll}
Channel, cadence & TDI $A$; $\Delta t=5$\,s \\
Samples & $N=1{,}572{,}864$ \\
Observation time & $\approx91.02$\,days \\
Free parameters & $\log_{10}A$, $\ln f_0$, $\ln\dot{f}_0$ \\
Fixed (rad) & $\iota=1.118$, $\phi_0=4.911$, $\psi=2.329$, \\
 & $\beta=0.981$, $\lambda=5.230$ \\
Prior & uniform, half-widths $(0.1,10^{-4},10^{-4})$ \\
Noise & stationary Gaussian, $A$-channel PSD \\
Gap masks & library of 1,000; 3.1--16.5\% missing \\
Clean waveforms & 24,000 training, 6,000 validation \\
Network input & $82\times768$ signed, noise-scaled WDM \\
\end{tabular}
\end{ruledtabular}
\end{table}

\subsection{Summarizer} \label{subsec:encoder}

The summarizer must keep the fine frequency structure of a slowly drifting track while relating parts of the observation that gaps separate. A convolutional stem with kernels elongated along time extracts local features and downsamples only along time, so that every frequency row keeps its resolution. At each downsampled time step the features of the whole band are projected to one token, which thus covers about one day of data. A learned class token and positional embeddings are added, and Transformer blocks~\cite{vaswani2017attention} let every token attend to every other one, so that information from segments on either side of a gap is combined directly rather than through the finite receptive field of a convolution. The context vector $c_\psi(\hat{d})$ combines the output class token with the mean of the other tokens; the mean keeps a direct average over all segments and reduces reliance on the single class token. The summarizer has $16.0\times10^6$ trainable parameters; Appendix~\ref{app:architecture} gives the layers.

\subsection{Flow matching} \label{subsec:cfm_engine}

FM learns a time-dependent velocity field whose flow transports a standard Gaussian at $t=0$ to the posterior at $t=1$~\cite{lipman2022flow, wildberger2023flow}. With parameters standardized by the training-set mean and standard deviation, $\boldsymbol{\theta}_1$, and a Gaussian draw $\boldsymbol{\theta}_0 \sim \mathcal{N}(0, I_3)$, we use the linear path $\boldsymbol{\theta}_t = (1 - t)\boldsymbol{\theta}_0 + t \boldsymbol{\theta}_1$, whose velocity is $\boldsymbol{\theta}_1 - \boldsymbol{\theta}_0$. The velocity network $v_\varphi$ and the summarizer are trained together by minimizing
\begin{equation} \label{eq:cfm_objective}
\mathcal{L}(\varphi,\psi) = \mathbb{E}_{t, \boldsymbol{\theta}_0, (\boldsymbol{\theta}_1, \hat{d})} \left\| v_\varphi\big(\boldsymbol{\theta}_t, t, c_\psi(\hat{d})\big) - (\boldsymbol{\theta}_1 - \boldsymbol{\theta}_0) \right\|^2,
\end{equation}
with $t \sim \mathcal{U}(0, 1)$ and $(\boldsymbol{\theta}_1, \hat{d})$ from the simulator. The objective is a regression and needs no ODE integration during training; its gradient reaches the summarizer through the context, which is thereby trained to keep what the posterior needs. The velocity network conditions each of its residual blocks on the context and on $t$ through adaptive layer normalization~\cite{perez2018film} ($7.5\times10^6$ parameters; the whole network has $23.5\times10^6$).

For an observation, the context is computed once; each posterior sample is obtained by drawing $\boldsymbol{\theta}_0$ and integrating $\mathrm{d}\boldsymbol{\theta}_t/\mathrm{d}t = v_\varphi(\boldsymbol{\theta}_t, t, c_\psi(\hat{d}))$ from $t=0$ to 1 with a fixed-step fourth-order Runge--Kutta scheme~\cite{hairer1993solving}. The log density of each sample, $\ln q(\boldsymbol{\theta}_1 \mid \hat{d})$, follows from that of $\boldsymbol{\theta}_0$ by integrating $\mathrm{d}\ln q_t(\boldsymbol{\theta}_t)/\mathrm{d}t = -\mathrm{Tr}\,[\partial v_\varphi/\partial \boldsymbol{\theta}_t]$ along the same path, where $q_t$ is the density of $\boldsymbol{\theta}_t$~\cite{chen2018neural, grathwohl2018ffjord}; with three parameters the trace is computed exactly. This density is what importance sampling requires (Sec.~\ref{sec:is}). We use FM because its regression objective trains the summarizer and the estimator together without constraining the architecture of the velocity network. The pipeline itself is general: any amortized posterior estimator that can be conditioned on the summary, such as a normalizing flow, could take the place of FM (Sec.~\ref{sec:discussion}).

\subsection{Training} \label{subsec:training_strategy}

Only the clean waveforms are stored. For every waveform in a mini-batch, a new noise realization and a mask from the library are drawn, and the observation is transformed and scaled [Eqs.~\eqref{eq:data_model}--\eqref{eq:noise_normalization}] on the GPU within the training step, so that repeated presentations of a waveform come with different noise and gaps and no image library is written to disk. The noise is thus new at every presentation, whereas the masks come from a finite set of 1,000 and the source parameters from the 24,000 training waveforms. The masks could also be generated on the fly, as the noise is; the library is cheaper, but fresh masks avoid a fixed set of gap patterns (Sec.~\ref{sec:discussion}). The network was trained with Adam~\cite{kingma2017adammethodstochasticoptimization} in \texttt{JAX}~\cite{jax2018github}, using \texttt{Equinox}~\cite{kidger2021equinox} and \texttt{Optax}~\cite{deepmind2020jax}, and the checkpoint with the lowest loss on the validation waveforms (with fixed noise and masks) was kept; training took about \TrainHours\,hours on one GPU.

\section{Results} \label{sec:results}

We evaluate the trained pipeline on new simulations in two ways. We first test whether its posteriors are calibrated (Sec.~\ref{subsec:calibration}), and then compare them with likelihood-based posterior estimates from PTMCMC (Secs.~\ref{subsec:setup} and~\ref{subsec:reference}).

\subsection{Calibration} \label{subsec:calibration}

\begin{figure*}[t]
\centering
\includegraphics[width=\textwidth]{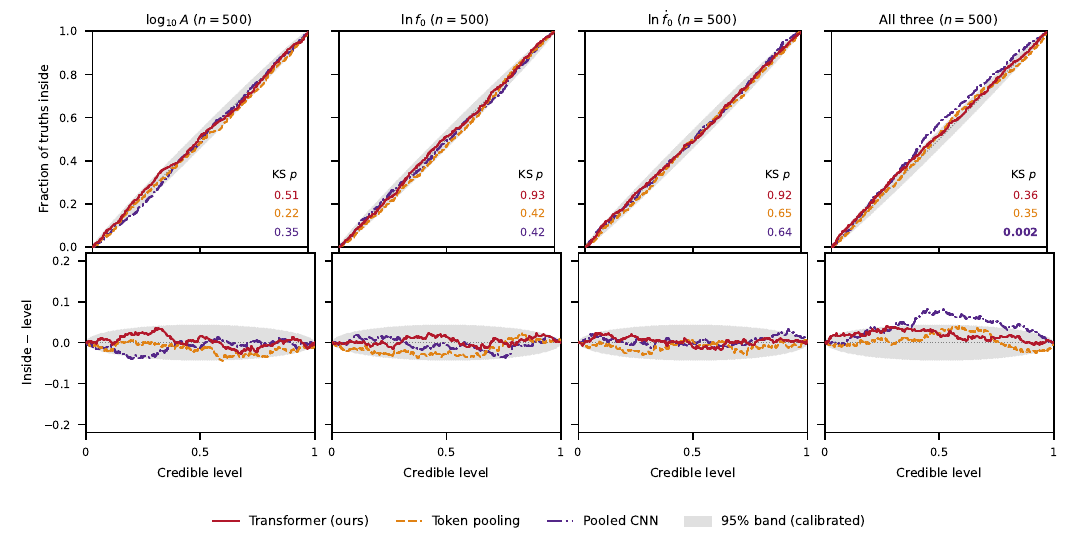}
\caption{Calibration over the \CalN{} new observations for the three summarizers. Top: coverage of credible regions for the marginal posteriors (first three panels from the left) and the joint posterior (rightmost panel); bottom: coverage error. Grey shaded area: 95\% band of the coverage expected for a calibrated posterior from \CalN{} observations; Numbers represent KS $p$-values (bold: $p<0.05$).}
\label{fig:calibration}
\end{figure*}

To assess the statistical reliability of the posteriors, we simulate \CalN{} new observations, each with parameters drawn from the prior, a new noise realization and gap pattern, not seen in training (Sec.~\ref{subsec:noise}; median missing fraction \CalMissing\%). Neither the signals nor the masks were used in training or in selecting the checkpoint. For each observation we draw 4,000 posterior samples and record the credible level at which the truth enters the posterior region: the central interval for one parameter, and the Gaussian ellipsoid of the posterior samples for several parameters jointly~\cite{hotelling1931generalization}. For a calibrated posterior these levels are uniformly distributed~\cite{talts2018validating}. Figure~\ref{fig:calibration} shows their P--P curves, with the Kolmogorov--Smirnov (KS) $p$-value of each, for our network and for two alternative networks that differ from it only in the summarizer (token pooling and a pooled CNN, Sec.~\ref{subsec:reference}). Our network passes all four tests: coverage of the 90\% credible region is about \CalCovRange\% and the smallest KS $p$ is \CalKSmin, as shown in Table~\ref{tab:calibration}. Token pooling also passes (smallest KS $p$ \CalKSTokMin). The pooled CNN passes for each parameter alone but fails for all three jointly (KS $p$ \CalKSCnnAll). It misrepresents the strong anticorrelation of $f_0$ and $\dot{f}_0$ (Table~\ref{tab:calibration}). A well-calibrated posterior is desirable; however, calibration of credible-region coverage alone does not establish the accuracy of the FMPD~\cite{Calibration_2019,PhysRevD.109.083002,hermans2022crisis}. A more direct comparison with PTMCMC is required.

\subsection{PTMCMC posteriors} \label{subsec:setup}

To compare with likelihood-based inference, we simulate \RefN{} observations in the same way. For each we sample the posterior with PTMCMC, using \texttt{ptemcee}~\cite{Vousden_2015, Foreman_Mackey_2013} and the WDM-domain likelihood~\cite{Digman_2023, pearson2025handling}
\begin{equation}\label{eq:wdm_likelihood}
\ln \mathcal{L}(\boldsymbol{\theta}) = -\frac{1}{2} \sum_{(n,m)\in\mathcal{B}} \frac{\left[ w_{nm}(\hat{d}) - w_{nm}(\hat{h}_{\boldsymbol{\theta}}) \right]^2}{S_m} + \text{const},
\end{equation}
where $\hat{h}_{\boldsymbol{\theta}} = m\odot h(\boldsymbol{\theta})$ is the template with the same mask as the data, $S_m$ is the variance of stationary-noise coefficients in row $m$, and $\mathcal{B}$ holds the 82 rows in $[1,5)$\,mHz that the network sees. All first runs use the same settings: \RefTemps{} temperatures with \RefWalkers{} walkers each, initialized uniformly in the prior, \RefSteps{} steps of which the first \RefBurn{} are discarded, leaving \RefRetained{} draws of the $T=1$ chain.

We assessed chain mixing and convergence of all \RefN{} PTMCMC runs using the potential scale reduction factor $\hat{R}$~\cite{gelman1992inference, brooks1998general, vehtari2021rank} and the effective sample size. \RefFirstPass{} of the \RefN{} first runs are converged. In the remaining \RefFailBadN{} runs, \RefStuckWalkers{} of the \RefWalkers{} walkers of the $T=1$ chain remain more than \RefStuckDist{} posterior standard deviations from the rest (Sec.~\ref{sec:is}), so these chains are not usable posteriors. We reran these \RefFailBadN{} injections with the same seed, \RefRerunTemps{} temperatures and \RefRerunSteps{} steps, and all \RefScreenPass{} PTMCMC posteriors reached a satisfactory level.

Equation~\eqref{eq:wdm_likelihood} neglects the noise correlations that gaps induce near their edges~\cite{Burke:2025, pearson2025handling}. PTMCMC therefore samples the posterior under a stated approximate likelihood, so we use it as a benchmark rather than as the exact posterior, whereas the network is trained on the full simulator.

\subsection{Comparison with PTMCMC} \label{subsec:reference}

\begin{figure*}[t]
\centering
\includegraphics[width=0.9\textwidth]{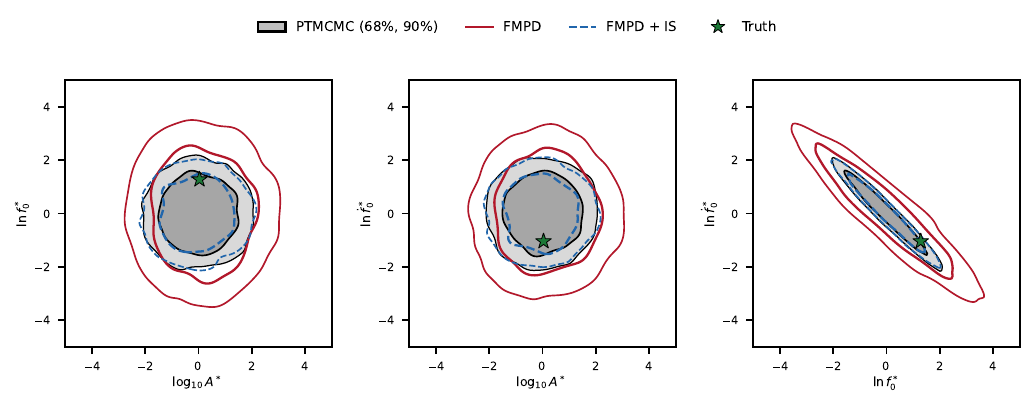}
\caption{Comparison of the PTMCMC posterior (filled, 68\% and 90\% regions), the FMPD and the FMPD corrected with IS (Sec.~\ref{sec:is}). The axes show $\theta^{*} = (\theta - \mu_{\rm PTMCMC})/\sigma_{\rm PTMCMC}$, each parameter standardized by the mean and standard deviation of the PTMCMC posterior.}\label{fig:example}
\end{figure*}

Figure~\ref{fig:example} shows one example, with \ExGap\% of the data missing. The FMPD is centred on the PTMCMC posterior and reproduces the strong anticorrelation of $\ln f_0$ and $\ln\dot{f}_0$, but it is wider.

\begin{figure*}[t]
\centering
\includegraphics[width=\textwidth]{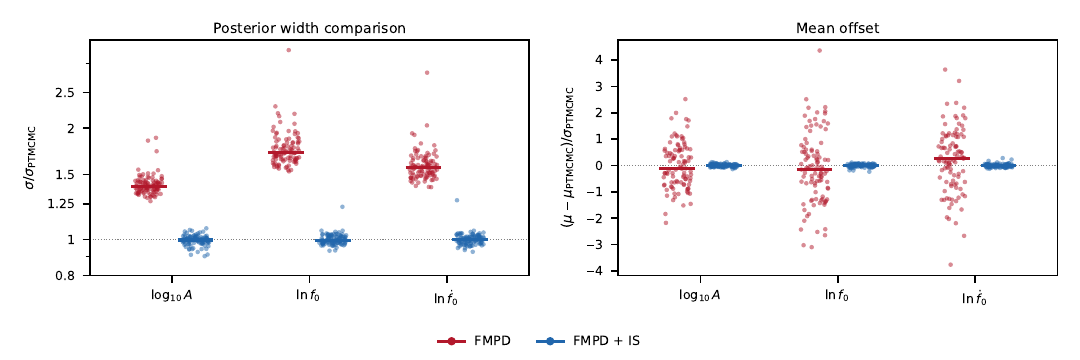}
\caption{The FMPD and FMPD + IS posteriors against PTMCMC posteriors over the \RefN{} new simulated observations. Left: posterior standard deviation relative to PTMCMC standard deviations; right: offset of the posterior mean in PTMCMC standard deviations.}
\label{fig:agreement}
\end{figure*}

Figure~\ref{fig:agreement} and Table~\ref{tab:agreement} compare the FM posteriors with PTMCMC over all 100 observations. The FM posteriors are wider than PTMCMC by median factors of \WidthRange, most for the two frequency parameters, which the phase evolution of the signal constrains, and their means are close to the PTMCMC means (median offset below one PTMCMC standard deviation). However, the FM posterior means lie further from the truth (by factors of \ErrRatio{} in root mean square, details in Appendix~\ref{app:truth}): the network extracts less information than the likelihood, and the comparison below suggests that most of this loss lies in the summary.

\begin{table*}
\caption{\label{tab:agreement} Agreements of FMPD and corrected FMPD with the PTMCMC posterior over the \RefN{} simulated observations. The first three rows differ only in the summarizer.}
\footnotesize
\begin{tabular}{lccccccc}
\hline\hline
 & \multicolumn{3}{c}{Width $\sigma/\sigma_{\rm PTMCMC}$\footnote{Median and 16--84\% range over all observations.}} & \multicolumn{3}{c}{Offset $|\mu-\mu_{\rm PTMCMC}|/\sigma_{\rm PTMCMC}$\footnote{Median over all observations.}} & \\
\cline{2-4}\cline{5-7}
Model & $\log_{10}A$ & $\ln f_0$ & $\ln\dot f_0$ & $\log_{10}A$ & $\ln f_0$ & $\ln\dot f_0$ & $D_{\rm KL}$ (nats)\footnote{Kullback--Leibler divergence of the Gaussian approximations of the joint posterior from the PTMCMC posterior; mean and 95\% bootstrap interval over all observations.} \\
\hline
Transformer (ours) & 1.39 (1.33--1.46) & 1.72 (1.62--1.89) & 1.57 (1.47--1.68) & 0.62 & 0.73 & 0.81 & 8.43 [7.38, 9.70] \\
Token pooling & 1.50 (1.44--1.59) & 1.77 (1.65--1.98) & 1.65 (1.56--1.83) & 0.74 & 1.11 & 1.03 & 12.08 [10.44, 13.85] \\
Pooled CNN & 1.55 (1.47--1.64) & 2.41 (2.24--2.75) & 2.17 (2.03--2.48) & 0.75 & 1.45 & 1.19 & 28.37 [23.62, 34.26] \\
\hline
FMPD + IS \footnote{Corrected FMPD (Sec.~\ref{sec:is}).} & 1.00 (0.97--1.02) & 0.99 (0.97--1.03) & 1.00 (0.97--1.03) & 0.04 & 0.04 & 0.04 & 0.012 [0.008, 0.017] \\
\hline\hline
\end{tabular}
\end{table*}

\textbf{Role of the summarizer.} Two networks that differ from ours only in the summarizer show how much of the difference from PTMCMC depends on it (Appendix~\ref{app:architecture}): \emph{token pooling} replaces the Transformer blocks by mean and max pooling over the tokens, and the \emph{pooled CNN} uses strided convolutions and global average pooling, without tokens. The mean Gaussian KL divergence from PTMCMC rises from \KLfm{} nats for the Transformer to \KLtok{} for token pooling and \KLcnn{} for the pooled CNN (Table~\ref{tab:agreement}), and the pooled CNN also misrepresents the correlation of $f_0$ and $\dot{f}_0$ (Sec.~\ref{subsec:calibration}). Attention lets the summarizer combine segments separated by gaps before pooling; part of the gain may also come from its larger size (16 million parameters, against 3.3 and 0.66 million).

\subsection{FMPD correction using importance sampling} \label{sec:is}

Where the likelihood can be evaluated, we use importance sampling (IS) to correct the FM posterior approximation $q$~\cite{dax2023neural}. With draws $\boldsymbol{\theta}_i \sim q(\boldsymbol{\theta} \mid \hat{d})$, $i=1,\dots,n$, the corrected posterior is
\begin{equation}\label{eq:is_weights}
\hat{q}(\boldsymbol{\theta} \mid \hat{d}) = \frac{\sum_{i=1}^{n} w_i\,\delta(\boldsymbol{\theta}-\boldsymbol{\theta}_i)}{\sum_{i=1}^{n} w_i}, \qquad w_i = \frac{\mathcal{L}(\boldsymbol{\theta}_i)\,p(\boldsymbol{\theta}_i)}{q(\boldsymbol{\theta}_i \mid \hat{d})},
\end{equation}
with $\delta$ the Dirac delta, so that $\hat{q}$ puts weight $w_i/\sum_j w_j$ on each draw $\boldsymbol{\theta}_i$, $\mathcal{L}$ of Eq.~\eqref{eq:wdm_likelihood}, the prior $p$, and $\ln q$ from the ODE (Sec.~\ref{subsec:cfm_engine}). As $n$ grows, $\hat{q}$ converges to the posterior under $\mathcal{L}$, the same target as PTMCMC, and the effective sample size of the weights measures how efficient the correction is. Figure~\ref{fig:agreement} and Table~\ref{tab:agreement} show that $q$ tends to be moderately overdispersed compared to the PTMCMC posterior, so it covers the high posterior probability region, which efficient importance sampling requires.

We drew 5,000 proposals for each of the \IsN{} observations. The median effective sample size is \IsESSmed{} (\IsEffMed\% of the draws). On all \RefN{} observations, the reweighted posteriors match PTMCMC posteriors (Figs.~\ref{fig:example} and~\ref{fig:agreement}): their mean Gaussian KL divergence from the PTMCMC posteriors, \KLis{} nats, is below that between the two halves of the PTMCMC chains themselves (\KLfloor).

Appendix~\ref{app:stuck} presents a few rare cases in which PTMCMC exhibited sampling instability, particularly when the true parameter values were close to the boundaries of the uniform prior: some walkers stayed far from the signal, trapped in a local maximum of the likelihood at the lower edge of the amplitude prior, since a template that misses the signal only adds power; running longer did not reliably free them. Unlike PTMCMC, the FMPD generates posterior samples directly without iterative exploration of the likelihood, thereby avoiding the associated chain-mixing difficulties: on all three injections its draws lie around the truth, as on the other injections.

Importance sampling corrects the FMPD towards the likelihood-based posterior targeted by PTMCMC by reweighting its samples with the likelihood of Eq.~\eqref{eq:wdm_likelihood}. Provided that $q$ adequately covers the target posterior, the weighted empirical distribution $\hat{q}$ converges to the posterior under this likelihood as the number of samples increases. However, this likelihood neglects the gap-induced noise correlations~\cite{Burke:2025, pearson2025handling}. Consequently, this target is an approximate posterior, rather than the exact posterior under the data-generating model. Misspecification of the covariance does not seem to introduce substantial bias in point estimation; however, as shown in Appendix~\ref{app:truth}, PTMCMC tends to underestimate uncertainty under the data-generating model, and so does the importance sampling-corrected FMPD. In contrast, the original FMPD, trained on the full simulator, remains well calibrated. Accounting for these correlations requires a more computationally demanding likelihood, involving the non-diagonal noise covariance of the observed data~\cite{Burke:2025}; given the length of the simulated observations ($N \approx 1.6\times10^6$), we therefore adopt the diagonal WDM likelihood of Eq.~\eqref{eq:wdm_likelihood}. With a likelihood that includes these correlations, the same procedure would converge to the exact posterior.

\subsection{Computing cost} \label{subsec:cost}
Table~\ref{tab:cost} compares the compute cost. The FMPD estimator draws its samples in \TimeFM\,s per simulated observation; importance sampling takes \TimeISwall\,s of wall time, against \TimePTwall\,s for a first PTMCMC run and \TimePTconvWall\,s per converged run. A converged PTMCMC run returns more effective samples, so the advantage of importance sampling here is reliability more than speed: it needs no burn-in or convergence diagnosis, its effective sample size flags a poor proposal at once, and it has no chain that can stick. Its own failure, a proposal that misses part of the posterior, is made less likely by the calibration of the FMPD.

\begin{table}
\caption{\label{tab:cost}Computing cost over 100 simulated observations: mean time to produce the posterior (sampling) and end to end with start-up and checks (wall), and median numbers of draws and of effective samples. Training is done once.}
\footnotesize
\setlength{\tabcolsep}{4pt}
\begin{tabular}{lrrrr}
\hline\hline
Training (once) \footnote{Training for the end-to-end amortized SBI (importance sampling correction adds no training).} & \multicolumn{4}{r}{28 h} \\
\hline
Per injection & Sampling & Wall & Draws & Effective \\
\hline
FMPD & 8.0 s & -- & 4,000 & 4,000 \\
FMPD + IS\footnote{Proposals with their densities plus the likelihood of each; Kish's effective sample size of the weights.} & 46 s & 48 s & 5,000 & 695 \\
PTMCMC & & & & \\
\quad first run\footnote{Including burn-in.} & 111 s & 129 s & 16,000 & -- \\
\quad per converged run\footnote{Total time of all runs, including the reruns of the 3 first runs that were not accepted, divided by the 100 accepted as converged; effective samples: smallest over the three parameters, from the autocorrelation of the chains.} & 147 s & 166 s & 16,000 & 2,643 \\
\hline\hline
\end{tabular}
\end{table}

\section{Discussion} \label{sec:discussion}

In this work, we presented an amortized SBI framework for parameter estimation from long-duration LISA data interrupted by scheduled and unscheduled gaps. Signed, noise-scaled WDM coefficients reduce a 3-month gapped time series to a time--frequency image in which each gap affects only the pixels that overlap it; a convolutional--Transformer summarizer compresses this image; and a flow-matching estimator conditioned on the summary draws posterior samples. The framework thereby handles the non-stationarity introduced by the gaps without imputation, windowing or the inversion of a dense noise covariance, and without evaluating a likelihood.

A primary contribution of this work is a summarizer that makes amortized inference computationally feasible for observations lasting months, which would otherwise impose substantial computational demands. Because it is trained jointly with the estimator, the compression is optimized for the information on the physical parameters rather than for a reconstruction of the data. The resulting posteriors are well calibrated over \CalN{} simulated observations whose waveforms, noise and masks were not seen in training. Our controlled comparison (Sec.~\ref{subsec:reference}) showed the importance of the summarizer architecture for posterior accuracy: with the same estimator, training data and schedule, replacing the Transformer blocks by token pooling, or the whole summarizer by a pooled convolutional network, moves the posteriors further from the PTMCMC benchmark, and the pooled network also misrepresents the strong anticorrelation of $f_0$ and $\dot f_0$. Self-attention lets the summarizer relate the segments on either side of a gap before the information is pooled; part of the gain may also come from the greater model capacity of the Transformer architecture. Moreover, the summarizer is not tied to FM or to this signal: it maps any gapped WDM image to a context vector, so the same architecture could condition other neural posterior or likelihood estimators, or be retrained for other LISA sources and gap models, although we have tested it only here.

We further assessed the amortized posteriors' agreement with likelihood-based inference (Sec.~\ref{subsec:reference}). The FM posteriors are approximately centred on the PTMCMC posteriors but exhibit moderately greater dispersion, with the FM posterior means lying further from the truth on average than the PTMCMC means (Appendix~\ref{app:truth}). In contrast, the PTMCMC posteriors underestimate uncertainty. These differences may reflect distinct sources of approximation error. The approximate likelihood used by PTMCMC may preserve sufficient information for accurate point estimation while underestimating uncertainty by neglecting gap-induced noise correlations. For FMPD, information loss during summarization or approximation error in the FM estimator may lead to less accurate posterior means and greater dispersion, while still providing approximately correct coverage. Finite-sample variability may also contribute to the observed differences. Where a likelihood can be evaluated, importance sampling can be added to correct the FMPD to the benchmark: reweighting it with the likelihood reproduces the posterior under that likelihood, here the target of PTMCMC, without simulating a Markov chain, also for the observations on which the first PTMCMC run stuck at the lower edge of the amplitude prior; here its advantage over PTMCMC lies more in reliability than in speed.

Amortization pays off where each likelihood-based analysis is expensive or must be repeated for many sources: an amortized estimator analyses an extreme mass-ratio inspiral in tens of seconds, where GPU-accelerated MCMC takes about two days~\cite{Liang_2025}; a global fit that catalogued about 12,000 Galactic binaries ran for about a week on four GPUs~\cite{Katz:2025erebor}; and a likelihood that includes the gap-induced noise correlations would make each PTMCMC run far more expensive (Sec.~\ref{sec:is}), whereas the network would cost the same.

Looking forward, several avenues exist to refine and extend this framework. First, the comparison of summarizers (Sec.~\ref{subsec:reference}) shows that the summarizer architecture has a substantial effect on the posteriors, which suggests that there is room for further improvement in summarizer design; for example, a summary that retains more of the phase evolution of the signal could bring the FM posterior closer to the exact posterior. Such a summary need not be learned by the network alone: physically motivated statistics, such as estimates of the frequency and its drift from the track in the WDM image or from a fast template fit, could be combined with the learned summary as further inputs to the estimator~\cite{Fearnhead2012, LucasMakinen_2025}. The kernel shapes of the convolutional stem and the time span of each token could also be related to the time--frequency resolution of the WDM transform and to the chirp of the target signals, and other time--frequency bases could be evaluated.

Second, more realistic and diverse gap patterns could be incorporated into the training simulations. The noise is already simulated afresh at every training step, but the masks are drawn from a library of 1,000 generated with one gap model, and the calibration masks, although not seen in training, come from the same model. Generating the masks on the fly, as the noise is, would avoid a fixed set of gap patterns and make it easy to train on other gap models. Likewise, the noise here is stationary apart from the gaps, whereas the noise level may change after a gap~\cite{pearson2025handling}; in future work we will include such changes in the simulator, since amortized SBI learns from simulations and needs no likelihood that models them.

Finally, the toy signal is a LISA-like chirping sinusoid with three free parameters within a local prior, not a realistic source model. It was chosen to show how the summary pipeline handles gaps, and to allow a comparison with PTMCMC over many injections. Future work will extend the framework to realistic source models, such as Galactic binaries with their full parameter space. The prior could also be widened by generating the waveforms during training, as is already done for the noise, instead of storing them, which limits the training set to 24,000 waveforms. Since the WDM transform separates signals in the time--frequency plane, the framework could then be extended towards the global fit of overlapping sources~\cite{littenberg:2023}: trained on multi-source realizations, the amortized posterior could slot into a global-fit pipeline as a fast proposal or initializer for individual sources, or, within an iterative block-wise scheme, as a dedicated update for one source class that acts on the residual data stream. Scaling amortized SBI to a full global fit must, however, overcome the dimensionality of joint forward modelling and an unknown number of overlapping sources, which calls for trans-dimensional network architectures. Moreover, since real LISA data contain no signal-free segments, the instrumental noise will have to be estimated jointly with the sources, including the row noise levels that scale the WDM coefficients [Eq.~\eqref{eq:noise_normalization}], which are estimated here from noise-only simulations.

The code is available on GitHub.\footnote{\url{https://github.com/bpandamao/WDM_CNN_TRANSFORMER_SBI}}

\section*{Acknowledgements}
We thank Avi Vajpeyi for suggesting the wavelet representation and for help with the \texttt{pywavelet} package~\cite{pywavelet, Cornish:2020:PhRvD, CornishWDMTransformRepo, MCDigmanWDMTransformRepo}. All computations were performed on a single NVIDIA L40S-48Q GPU under Ubuntu Linux; the networks were implemented in \texttt{JAX}. We thank Associate Professor Simon Urbanek, the Department of Statistics, and the Science Connect IT team at the University of Auckland for access to and assistance with the High Performance Computing facilities. Ruiting Mao thanks the University of Auckland for a UoA Doctoral Scholarship. Matthew C. Edwards and Jeong Eun Lee acknowledge support by the Marsden grant MFP-UOA2531 from New Zealand Government funding, administered by the Royal Society Te Ap\={a}rangi.

\appendix \label{app}

\section{Network architecture} \label{app:architecture}

\subsection{Parameter count breakdown}
Table~\ref{tab:parameter_counts} reports the trainable parameter counts of each component, based on the 3-month toy signal. The entire model comprises $23{,}476{,}195$ parameters, with the summarizer accounting for $\approx 68\%$ ($15{,}978{,}976$) and the flow-matching velocity network for $\approx 32\%$ ($7{,}497{,}219$).

\begin{table*}[t]
\centering
\caption{Trainable parameter breakdown of the inference pipeline.}
\label{tab:parameter_counts}
\begin{ruledtabular}
\begin{tabular}{lr}
\textbf{Sub-module} & \textbf{Parameters} \\
\hline
\multicolumn{2}{l}{\textbf{Summarizer}} \\
\quad Asymmetric 2D CNN stem (3 Conv2d + GroupNorm) & $113{,}120$ \\
\quad Token projection ($5248 \to 512$) + LayerNorm & $2{,}688{,}512$ \\
\quad Class token + positional embeddings ($97\times512$) & $50{,}176$ \\
\quad Transformer blocks ($4\times$, attention + MLP) & $12{,}601{,}344$ \\
\quad Hybrid pooling (LayerNorm + Linear $1024 \to 512$) & $525{,}824$ \\
\quad \textit{Total summarizer} & $\mathbf{15{,}978{,}976}$ \\
\hline
\multicolumn{2}{l}{\textbf{Conditional flow matching}} \\
\quad Time-embedding MLP ($256\to512\to512\to512$) & $656{,}896$ \\
\quad Conditioning fusion (Linear $1024 \to 512$) & $524{,}800$ \\
\quad State lifting (Linear $3 \to 512$) & $2{,}048$ \\
\quad AdaLN residual blocks ($4\times$) & $6{,}311{,}936$ \\
\quad Velocity readout (Linear $512 \to 3$) & $1{,}539$ \\
\quad \textit{Total velocity network} & $\mathbf{7{,}497{,}219}$ \\
\hline
\textbf{Total end-to-end} & $\mathbf{23{,}476{,}195}$ \\
\end{tabular}
\end{ruledtabular}
\end{table*}

\subsection{Asymmetric 2D convolutional stem}
The noise-scaled WDM image $U \in \mathbb{R}^{1 \times 82 \times 768}$ is processed by three convolutional stages (Table~\ref{tab:cnn_stem}). Each stage applies a 2D convolution with a kernel elongated along the time axis and stride $(1, 2)$, so that only the time dimension is downsampled, followed by Group Normalization and a GELU activation.

\begin{table*}[t]
\centering
\caption{Asymmetric 2D convolutional stem. Kernels are given as (frequency, time) and outputs as (channels, frequency, time). Each stage uses stride $(1, 2)$, zero padding $(k-1)/2$ along each axis, GroupNorm with $C/4$ groups, and GELU. Parameter counts include convolution weights and biases and the GroupNorm affine terms.}
\label{tab:cnn_stem}
\begin{ruledtabular}
\begin{tabular}{ccccr}
Stage & Channels & Kernel & Output & Parameters \\
\hline
1 & $1 \to 16$ & $(7, 15)$ & $(16, 82, 384)$ & $1{,}728$ \\
2 & $16 \to 32$ & $(7, 11)$ & $(32, 82, 192)$ & $39{,}520$ \\
3 & $32 \to 64$ & $(5, 7)$ & $(64, 82, 96)$ & $71{,}872$ \\
\hline
\multicolumn{4}{l}{Total} & $113{,}120$ \\
\end{tabular}
\end{ruledtabular}
\end{table*}

\subsection{Transformer summarizer specifications}
At each of the $n = 96$ time steps, the feature column of size $C_3 \times F = 64 \times 82 = 5{,}248$ is projected to dimension $d = 512$ via $\text{Linear}(5248 \to 512)$ followed by $\text{LayerNorm}(512)$. A learnable [CLS] token, initialized as $e_{\text{CLS}} \sim \mathcal{N}(0, 0.02^2 I)$, is prepended, and learnable 1D positional embeddings $E_{\text{pos}} \in \mathbb{R}^{97 \times 512}$ are added.

The sequence is routed through $K = 4$ pre-LN Transformer layers:
\begin{itemize}
    \item Multi-Head Self-Attention ($\text{MHSA}$): model dimension $d = 512$, $H = 8$ attention heads, head dimension $d_k = 64$, without bias terms in the attention projections.
    \item Feed-Forward Network ($\text{FFN}$): two-layer MLP with expansion factor 4 ($\text{Linear}(512 \to 2048) \to \text{GELU} \to \text{Linear}(2048 \to 512)$).
    \item Residual skip connections applied across both attention and feed-forward sub-layers.
\end{itemize}
Following the final layer normalization $\tilde{Z} = \text{LayerNorm}(Z_4)$, hybrid pooling extracts
\begin{equation}
c_{\text{CLS}} = \tilde{Z}[0], \qquad c_{\text{mean}} = \frac{1}{96}\sum_{i=1}^{96}\tilde{Z}[i],
\end{equation}
which are concatenated into $\mathbb{R}^{1024}$ and projected to the final context vector
\begin{equation}
c_\psi(\hat{d}) = \text{GELU}\left( W_c [c_{\text{CLS}} \,\|\, c_{\text{mean}}] + b_c \right) \in \mathbb{R}^{512}.
\end{equation}

\subsection{AdaLN / FiLM residual blocks of the velocity network}
The flow time $t$ is mapped to a 256-dimensional sinusoidal embedding with 128 frequencies and passed through an MLP with two hidden layers of width 512 and SiLU activations, producing $t_{\text{emb}} \in \mathbb{R}^{512}$. The conditioning representation $z \in \mathbb{R}^{512}$ is constructed by concatenating the context $c_\psi(\hat{d})$ with $t_{\text{emb}}$:
\begin{equation}
z = \text{SiLU}\left( W_f [c_\psi(\hat{d}) \,\|\, t_{\text{emb}}] + b_f \right).
\end{equation}
The parameter state is lifted as $h_0 = \text{SiLU}(W_{\text{in}}\boldsymbol{\theta}_t + b_{\text{in}}) \in \mathbb{R}^{512}$. Within each of the $M = 4$ residual blocks, the conditioning vector $z$ is linearly mapped to four modulation vectors,
\begin{equation}
[\gamma_1, \beta_1, \gamma_2, \beta_2] = \text{Linear}(512 \to 2048)(z),
\end{equation}
where each $\gamma_k, \beta_k \in \mathbb{R}^{512}$. Each residual block executes
\begin{align}
h_1 &= \text{SiLU}\left( W_1 \left( \text{LN}(h) \odot [1 + \gamma_1] + \beta_1 \right) + b_1 \right), \\
h_2 &= W_2\, \text{SiLU}\left( \text{LN}(h_1) \odot [1 + \gamma_2] + \beta_2 \right) + b_2, \\
h_{\text{out}} &= h + h_2,
\end{align}
where $W_1, W_2 \in \mathbb{R}^{512 \times 512}$. A final projection $\text{Linear}(512 \to 3)$ maps $h_{\text{out}}$ of the fourth block to the estimated vector field $v_\varphi(\boldsymbol{\theta}_t, t, c_\psi(\hat{d})) \in \mathbb{R}^3$.

\subsection{Summarizers of the comparison}
The two networks of Sec.~\ref{subsec:reference} use the same flow-matching network, training data and schedule. \emph{Token pooling} keeps the convolutional stem and token projection and replaces the Transformer blocks and class token by the concatenated mean and maximum over the tokens; its summarizer has $3{,}327{,}456$ parameters ($10{,}824{,}675$ in total). The \emph{pooled CNN} applies five strided convolutional stages followed by global average pooling over frequency and time, without tokens; its summarizer has $663{,}648$ parameters ($8{,}160{,}867$ in total).

\section{Numerical precision}  \label{app:training_reproducibility}

Detector noise realizations are generated in the frequency domain using the analytic LISA noise PSD $S_n(f)$. The Fourier noise standard deviation $[N S_n(f) / (4\Delta t)]^{1/2}$ is computed in 64-bit double precision before being cast to 32-bit single precision at the network boundary, because the variance itself can underflow in single precision for strain in physical units. The moments used to estimate the row scales $\sigma_f$ of Eq.~\eqref{eq:noise_normalization} are likewise accumulated in double precision.

\section{Calibration table} \label{app:calibration}

Table~\ref{tab:calibration} lists the KS $p$-values and 90\% coverages behind Fig.~\ref{fig:calibration} for the three summarizers on the calibration set; it also adds the joint test of $(\ln f_0, \ln\dot{f}_0)$.

\begin{table*}
\caption{\label{tab:calibration}Calibration over the 500 new simulated observations: KS $p$-value of the credible level at which the truth enters the posterior region (bold: $p<0.05$), and the percentage of truths inside the 90\% region (central interval for one parameter, Gaussian ellipsoid for two or three).}
\footnotesize
\begin{tabular}{lccccc}
\hline\hline
 & $\log_{10}A$ & $\ln f_0$ & $\ln\dot f_0$ & $(\ln f_0, \ln\dot f_0)$ & All three \\
\hline
\multicolumn{6}{l}{\emph{KS $p$}} \\
Transformer (ours) & 0.51 & 0.93 & 0.92 & 0.29 & 0.36 \\
Token pooling & 0.22 & 0.42 & 0.65 & 0.34 & 0.35 \\
Pooled CNN & 0.35 & 0.42 & 0.64 & \textbf{$<$0.001} & \textbf{0.002} \\
\multicolumn{6}{l}{\emph{Truths inside the 90\% region (\%)}} \\
Transformer (ours) & 91 & 92 & 91 & 93 & 90 \\
Token pooling & 88 & 91 & 89 & 90 & 88 \\
Pooled CNN & 90 & 90 & 92 & 95 & 93 \\
\hline\hline
\end{tabular}
\end{table*}

\section{Calibration of the PTMCMC posteriors} \label{app:truth}
\begin{figure*}[t]
\centering
\includegraphics[width=\textwidth]{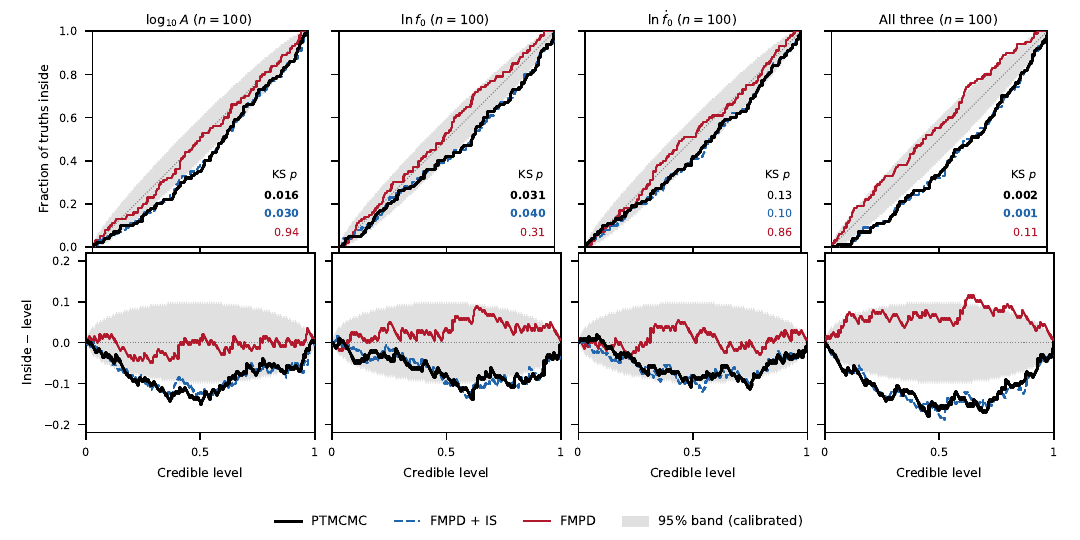}
\caption{Calibration with respect to the injected values over the \RefN{} new simulations, shown as in Fig.~\ref{fig:calibration}, for PTMCMC, FMPD + IS and the FMPD. Below the diagonal (top) or below zero (bottom), fewer truths lie inside the regions than their credible level, so the posterior is over-confident.}
\label{fig:truth}
\end{figure*}

Figure~\ref{fig:truth} compares the calibration of the PTMCMC, FMPD + IS, and FMPD posteriors against the injected parameter values across \RefN{} new simulated observations, using the same diagnostics as Fig.~\ref{fig:calibration}. The nearly identical coverage curves of PTMCMC and FMPD + IS indicate that importance sampling reproduces the calibration properties of the PTMCMC posterior, consistent with their close agreement in Table~\ref{tab:agreement}. However, both are over-confident: their curves lie below the diagonal, and the 90\% regions of PTMCMC contain the truth in \TruthRefCov\% of the injections rather than 90\%.

Importance sampling uses no Markov chain, so the over-confidence points to the likelihood of Eq.~\eqref{eq:wdm_likelihood} rather than the sampler; one possible cause is the noise correlations that this likelihood neglects near the gaps. The FM posterior, trained on the full simulator, passes all four tests and is slightly conservative in the joint space. The standardized errors $(\theta_{\rm true}-\bar\theta)/\sigma$ of PTMCMC have root mean squares of \TruthZRef{}, which exceed the theoretical benchmark of 1 under a correctly specified posterior, so its posteriors are slightly narrow by about this factor. This accounts for only part of the factors of \WidthRange{} by which the FMPD is wider than PTMCMC (Sec.~\ref{subsec:reference}); the rest reflects a loss of information, since the FMPD means also lie further from the truth, by factors of \ErrRatio{} in root mean square. Since the injections are drawn from the prior, no estimate has a smaller expected squared error than the mean of the exact posterior, so an FMPD mean with a larger error than the PTMCMC mean shows that the FMPD uses less information than the data contain, whatever the target of PTMCMC.

\section{PTMCMC runs that stuck} \label{app:stuck}
\begin{figure*}[t]
\centering
\includegraphics[width=\textwidth]{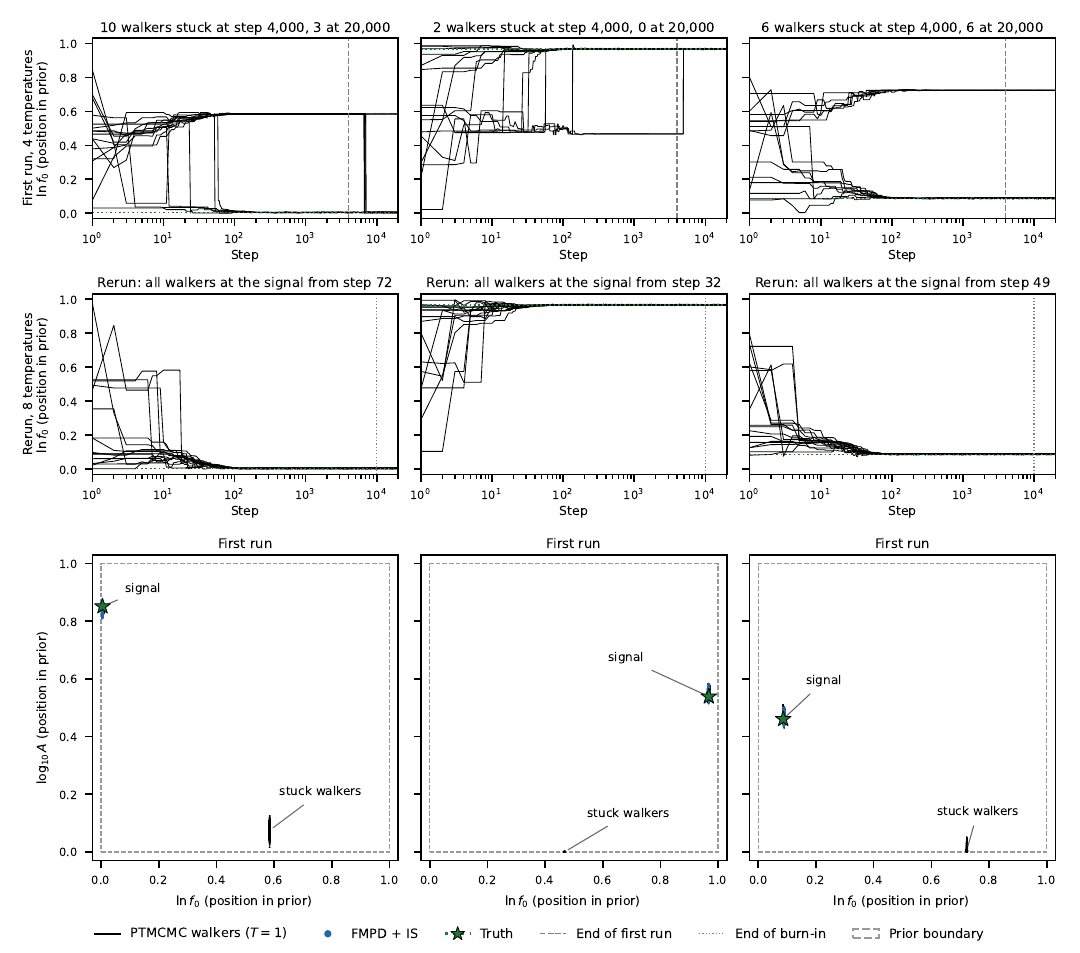}
\caption{The \RefFailBadN{} injections whose first PTMCMC run stuck, one column each. Top: $\ln f_0$ of the $T=1$ walkers in the first run, continued with the same seed and ladder; the first run ends at the dashed line. Middle: the same in the rerun used as the reference; dotted: end of its burn-in. Both on a logarithmic step axis. Bottom: the retained draws of the first run and the FMPD + IS posterior over the whole prior in the $(\ln f_0, \log_{10}A)$ plane, from 0 at the lower to 1 at the upper prior bound; the posterior around the signal lies under the star, and the stuck walkers appear as a dot or a short vertical line depending on how far they spread in amplitude.}
\label{fig:stuck}
\end{figure*}

Figure~\ref{fig:stuck} shows the \RefFailBadN{} injections whose first PTMCMC run stuck (Sec.~\ref{sec:is}). The stuck walkers sit near the lower amplitude bound, far from the signal, whose posterior, from PTMCMC or from FMPD + IS, is a point under the star at the scale of the prior (bottom). In all three, the injected $f_0$ lies close to a bound of its prior (within 9\% of the prior range), as it does for only 15 of the other 97 injections. A signal near the edge of the prior may be harder for the walkers to reach, although most injections with $f_0$ this close to a bound did not stick.

Continuing the first runs with the same ladder to \RefContSteps{} steps (top) freed some of the stuck walkers in the first injection (left), all of them in the second and none in the third. The walkers left only by swaps with a hotter chain, and almost all walkers of the hotter chains sat at the same low-amplitude edge. We reran these injections with the same seed and starting positions but \RefRerunTemps{} temperatures (middle), and all three reruns converged satisfactorily.

\bibliographystyle{apsrev4-1}
\bibliography{refs}

\end{document}